\documentclass[12pt,preprint]{aastex}
\shorttitle{Galactic Metallicity Gradients}
\shortauthors{Daflon \& Cunha}
\begin{document}

\title{Galactic Metallicity Gradients Derived from a Sample of OB Stars} 

\author{Simone Daflon}
\affil{Observat\'orio Nacional, Rua General Jos\'e Cristino 77 \\
CEP 20921-400, Rio de Janeiro  Brazil }
\email{daflon@on.br}
\author{Katia Cunha}
\affil{ Observat\'orio Nacional, Rua General Jos\'e Cristino 77 \\
 CEP 20921-400, Rio de Janeiro  Brazil \\
 and \\
Department of Physics, University of Texas at El Paso \\
 El Paso, TX 79968-0515}
\email{katia@baade.physics.utep.edu}

\clearpage
\begin{abstract}

The distribution of stellar abundances along the Galactic disk
is an important constraint for models of chemical evolution and
Galaxy formation. In this study we derive radial gradients of 
C, N, O, Mg, Al, Si, as well as S, from abundance
determinations in young OB stars.
Our database is composed of a sample of 69 members
of 25 open clusters, OB associations and H~II regions with 
Galactocentric distances between 4.7 and 13.2 kpc. 
An important feature of this abundance database is the fact that
the abundances were derived self-consistently in non-LTE using
a homogeneous set of stellar parameters.
Such an uniform analysis is expected to reduce the magnitude
of random errors, as well as the influence of systematics in the 
gradients defined by the abundance and Galactocentric distance. 
The metallicity gradients obtained in this study are, in general, flatter 
than the results from previous recent abundance studies of early-type stars. 
The  slopes are found to be between $-0.031$ (for oxygen) and 
$-0.052 {\rm \,dex \, kpc^{-1}}$ (for magnesium).
The gradients obtained for the studied elements are quite similar and if
averaged, they can be represented by a single 
slope of $-0.042 \pm 0.007 {\rm \,dex \, kpc^{-1}}$.
 This value is generally consistent with 
an overall flattening of the radial gradients with time.
\end{abstract}


\newpage

\section{Introduction}\label{intr}

Chemical evolution models of the Galaxy must reproduce certain
observational constraints, such as the age-metallicity relation, the
abundance patterns of different stellar populations and the chemical composition
of the Sun. Reliable observational data for a variety of objects
(where observational data in this case means results from abundance analyses)
are thus crucial in constraining
the assumptions that enter in the construction of such models.
One very important observational model constraint is the radial metallicity
gradient, or,
the distribution of abundances in the Galactic disk as a function
of Galactocentric distance, $R_g$. In general terms, abundance gradients
are a feature commonly observed in all galaxies with their metallicities
decreasing outwards from the galactic centers.
Metallicity gradients appear to be shallower in elliptical galaxies and progressively
increase from lenticulars to barred spirals, being steeper
in normal spirals.  (See e.g. the review by Henry \& Worthey 1999).
For these extragalactic systems the metallicity is traced via analyses
of individual H~II regions \citep{vee92,ken03} and by the most luminous stars in galaxies of 
the local group (e.g. Monteverde et al. 1997), or by photometric properties.
In our own Galaxy, metallicity gradients can be determined via
abundance analyses of ionized gas in H~II regions and
planetary nebulae, chemical compositions of stellar photospheres of young OB stars 
and the more evolved Cepheids, or from studies of metallicities of open clusters.  
Although significant efforts have been put into
trying to pin down metallicity gradients for the Galactic disk, the abundance slopes
for the different elements in these different populations remain somewhat uncertain.

Focusing first on metallicity gradients inferred from analyses of
young OB stars, there seems to be no consensus between the different determinations
published over the last 10--15 years.  The first studies of
\citet{fiz90} suggested  almost null gradients of oxygen and magnesium,
while for N, Al and Si they found a tendency of increasing
abundances with Galactocentric radius. Later on,
\citet{kau94} combined their own abundance determinations of N and O with those
from the literature and obtained a zero gradient for oxygen and
$ -0.026 {\rm \,dex\,kpc^{-1}}$ for nitrogen. \citet{km94} derived
abundance gradients for several elements (C, N, O, Mg, Al, Si in non-LTE and
Ne, S, Fe in LTE)
that were in general flatter than $-0.03 {\rm \,dex\,kpc^{-1}}$
in the range between $R_g$= 5--15 kpc. (Moreover, their results suggested that the abundance
distribution in the local ISM is inhomogeneous.)
More recent studies, however, 
obtained considerably steeper gradients for the Galactic disk.
The self-consistent non-LTE study of 16 target stars by
\citet{gum98} resulted in a gradient of
$-0.067 {\rm \,dex\,kpc^{-1}}$ for oxygen; for other elements (such as
C, N, Mg, Al, and Si), the gradients varied between
$-0.035 {\rm \,{dex\, kpc^{-1}}}$ (for carbon) and
$-0.107 {\rm \,{dex\, kpc^{-1}}}$ (for silicon).
\citet{ser97} compiled data for   early-type stars from  
abundance papers previously published by their group  and obtained 
an oxygen gradient of  $-0.07 {\rm \,dex\,kpc^{-1}}$. 
Other elements like  C, N, Mg, Al, and Si, and oxygen included, were analyzed
by \citet{rol00}, who found gradients that are steep, but are all similar in 
magnitude, with a mean value of $-0.068 {\rm \,dex\,kpc^{-1}}$. 
This stellar database was later extended towards the innermost disk by \citet{sma01}, 
providing gradients consistent with those derived by \citet{rol00} for the disk, 
except for oxygen.

Several studies of the Milky Way metallicity gradients are based on chemical analyses of
ionized gas in H~II regions and planetary nebulae.
Oxygen abundances for a sample of 21 H~II regions between 5.9 and
13.7 kpc by \citet{sha83}
indicate a radial gradient of $-0.07 {\rm\, dex\,kpc^{-1}}$.
Other nebular studies concentrate on
specific parts of the Galactic disk. In the outer disk, for example,
\citet{vee96} suggest that the N, O, and S gradients are much flatter
-- when compared to \citet{sha83} -- in the direction of Galactic anti-center.
\citet{aff97} studied
the region inner to $R_g \sim 11.4$ kpc and obtained
gradients for N, O, and S, of about  $-0.07 {\rm\, dex\,kpc^{-1}}$.
However, the most recent nebular studies tend to
find flatter gradients for oxygen.
\citet{deh00} obtained a gradient of $-0.039  {\rm\, dex\,kpc^{-1}}$ for
oxygen in H II regions between $R_g$= 5--15 kpc.  
\citet{pil03} did a compilation of spectra from  H~II regions located between
6.6 and 14.8 kpc and 
obtained a gradient of $-0.048 {\rm\, dex\,kpc^{-1}}$.
The planetary nebulae (PNe) ejected from low- to intermediate-mass stars
can also trace the abundance gradients. 
The study of \citet{meq99}, based on abundance data compiled from the literature,
found gradients of $-0.058 {\rm\, dex\,kpc^{-1}}$ for oxygen
and  $-0.077 {\rm\, dex\,kpc^{-1}}$ for sulfur.
\citet{mev00} recomputed temperatures and abundances from line ratios
published in the literature and obtained
$-0.054$ and $-0.064 {\rm\, dex\,kpc^{-1}}$ for
oxygen and sulfur, respectively. 
More recently, however,  
the analysis of PNe by \citet{hkb04} derived flatter gradients 
of $-0.037\pm0.008$ and $-0.048\pm0.0098 {\rm\, dex\,kpc^{-1}}$ for O and S, 
respectively.

In fact, the actual picture seems to be more complex as studies of 
metallicities in open clusters with different ages  \citep{fri02,chw03} 
and abundances of planetary nebulae \citep{mcu03} suggest that the  
metallicity gradients may evolve, showing a tendency of becoming
flatter with time. Besides, there may be 
a discontinuity in the radial Galactic gradient with two distinct metallicity
zones in the disk (results from open clusters from Twarog, Ashman, \& Anthony-Twarog
1997 and Cepheids by Caputo et al. 2001; Andrievsky et al. 2004).
Concerning the population of young OB stars, a determination of radial
gradients that is  based on a homogeneous and self-consistent non-LTE abundance
database would represent a significant advancement. This is the goal of this paper: 
the last 
in a series. In this study we assemble our previous results (Paper I - VI;
where we progressively built a non-LTE abundance database of OB stars from
high-resolution spectra), in order to better define the current metallicity gradients
of the Milky Way disk.

\section{The Abundance Database}\label{abund}

Our observational data consisted of 35 high resolution echelle spectra obtained
with the 2.1m telescope at the McDonald Observatory (University of Texas, Austin)
and 34 obtained with the 1.52m telescope at the European Southern Observatory (La Silla, Chile).
Our Northern sample was composed basically of stellar members of nearby OB
associations, within 2.5 kpc from the Sun, while the
sample stars observed from the South were distributed along the Galactic
disk ranging from 4.7 to 13.2 kpc in Galactocentric distance.

Our database contains non-LTE abundances of C, N, O, Mg, Al, Si, and S
for 69 late-O to early-B type star members of
25 OB associations, open clusters or H II regions (Papers I$-$VI), plus
18 stellar members of the Orion association from \citet{cel94}.
All these abundances were derived with the same methodology and are
strictly on the same scale. We observed several stars per cluster or
association in order to
sample, although to a very modest degree, a cluster's metallicity distribution.
Our observing campaigns included 108 additional stars that
had to be discarded due to their large projected rotational velocities
(the distributions of $v \sin i$ values for our whole sample of stars members
of associations will be the subject of a future study), or because they
were binary. The association of Cep~OB2 was the most extensively probed,
with 17 targets analyzed for abundances; for 14 other clusters
we had five or less sample stars while for ten others we had to rely on
the abundance derived for one target star in order to represent the cluster
abundance. Average abundances for each cluster are listed in
Table~\ref{tab1}, together with the number [n] of stars analyzed in each cluster.

\section{The Adopted Distances}\label{dist}

Galactocentric distances are a crucial ingredient in establishing radial metallicity gradients.
Table~\ref{tab2} lists our target clusters, their Galactic coordinates (columns 2 and
3), and their distances from the Sun (column 4), which were collected
from different studies in the literature. 
Most of these distances were derived from color-magnitude diagrams or from spectral
type {\it vs.} intrinsic color calibrations. In addition, distances  from  HIPPARCOS
parallaxes are available for our nearest target stars that belong to
the Ori OB1, Lac OB1 and Cep OB2 associations \citep{dez99}.
The typical errors of the distances from the Sun are inferred to be of the
order of 10--20\% in the individual studies, but can reach 40\% in a few cases.
An inspection of different literature values for a given cluster shows that
the different distance estimates for some clusters are quite discrepant. 
The largest dispersions (sigmas between 20-50 percent)
are found for NGC~6204, NGC~6604, Vul~OB1,
NGC~4755, Cep~OB2, Sh2~247, NGC~1893 and Sh2~285. The latter, being the furthest H~II
region in our sample, deserves special mention, as it plays an important role in
the definition of the abundance gradient.
\citet{lah87} considered 5 stars in this HII region, and obtained a
spectroscopic distance of 6.4 kpc.
\citet{mof79} derived d=6.9 kpc based on
UBV photometry and spectral types of two stars; the same distance was
obtained by \citet{tem93} from the fitting of theoretical isochrones for
eight stars. However, \citet{rol94} derived a smaller distance of d=4.3 kpc,
based on theoretical evolutionary tracks for two stars in Sh2~285.
Here, we adopt the average distance of $5.9\pm1.4$ kpc for Sh2~285.
For each target cluster we calculated average distances and
dispersions from all the distance estimates. These average distances
are listed in column 5 of Table~\ref{tab2}. 

With the cluster distances from the Sun, d,
Galactocentric distances projected onto the Galactic plane can be
calculated from
$$ {R^2_g} = {R^2_\odot} +  (d \cos b)^2 - 2 R_\odot d \cos l \cos b, $$
where  ${R_\odot}$ is the Galactocentric distance of the Sun  
and  $l,b$ are the Galactic
longitude and latitude of the object, respectively.
Here, the distance of the Sun from the Galactic center is
taken to be 7.9 kpc (MacNamara et al. 2000; this value
agrees well with the weighted average previously calculated by Reid 1993
computed from published values derived from various approaches,
R$_\odot=8.0\pm0.5$~kpc).

\begin{figure}
\epsscale{0.75}
\plotone{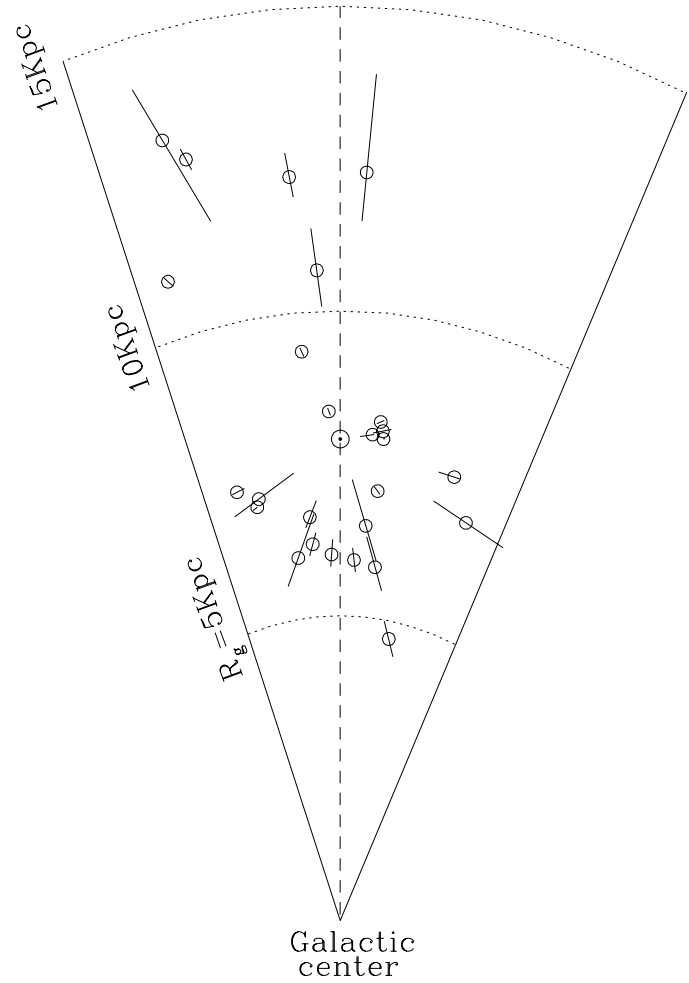}
\caption{The distribution of target clusters and associations projected onto a section 
of the Galactic Plane. The Sun is represented at $R_{\odot}$=7.9 kpc \citep{mac00}. 
The dashed line connects the Sun to the Galactic center and represents 
the direction $l=0,180^\circ$; concentric circles at the distances of 
5, 10 and 15 kpc of the Galactic center are depicted by dotted lines. 
The inclined bars show the range of $R_g$ 
calculated from different distances in the literature for a given cluster position.   
\label{fig1}}
\end{figure}

The derived cluster Galactocentric distances and uncertainties are listed in
column 6 of Table~\ref{tab2}; these will be adopted in the gradient calculations.
The errors in $R_g$ were computed by noting that these distances
depend upon the solar Galactocentric distance, the distance from the
Sun to the cluster (d), and the cluster Galactic latitude and longitude.
The total uncertainty then consists of the quadratic sum of the products of
each of these quantities by the corresponding uncertainties.
In Figure~\ref{fig1} we show the space distribution of our sample and the
positions of the sample clusters (open circles) projected onto a section of the
Galactic plane. The range in Galactocentric distance spanned by the
different determinations in the literature for each cluster are represented
by the bars in each location (the bar sizes are set by the smallest and 
largest distances in each case). 
Our database samples the inner arm of Sagittarius-Carina
(objects within $295^\circ < l < 360^\circ$ and $360^\circ < l < 18^\circ$),
the extension of the Perseus arm  ($174^\circ < l < 214^\circ$, with distances
larger than 1.5 kpc), as well as the local Orion spur ($40^\circ < l <206^\circ$),
where the Sun lies (Han et al. 2002, after Georgelin \& Georgelin 1976).

\section{The Radial Gradients}\label{grads}
 
One of the simplest ways to represent the distribution of radial metallicities across 
the Galactic disk is to assume that abundance trends  can be represented
by  straight lines spanning the whole interval in Galactocentric distance covered 
by the sample.
Radial metallicity gradients can then be obtained from linear fits of the form  
$a + bx$ to the abundances as a function of Galactocentric distances; the
abundance gradient, therefore, being represented by the single coefficient $b$. 
From our database of C, N, O, Mg, Al, Si and S abundances for clusters and
associations (Table~\ref{tab1}) and their Galactocentric distances (Table~\ref{tab2}),
best-fit straight lines ($a\pm\sigma a$ and $b\pm\sigma b$)
and correlation coefficients (R) were calculated. These are listed in 
Table~\ref{tab3} together with the number, $n$, of clusters or associations considered for 
each element. Here, the gradients were calculated assuming that the cluster abundances are
represented by {\it averages} of the abundances obtained for their sample stars. 

\begin{figure}
\epsscale{0.75}
\plotone{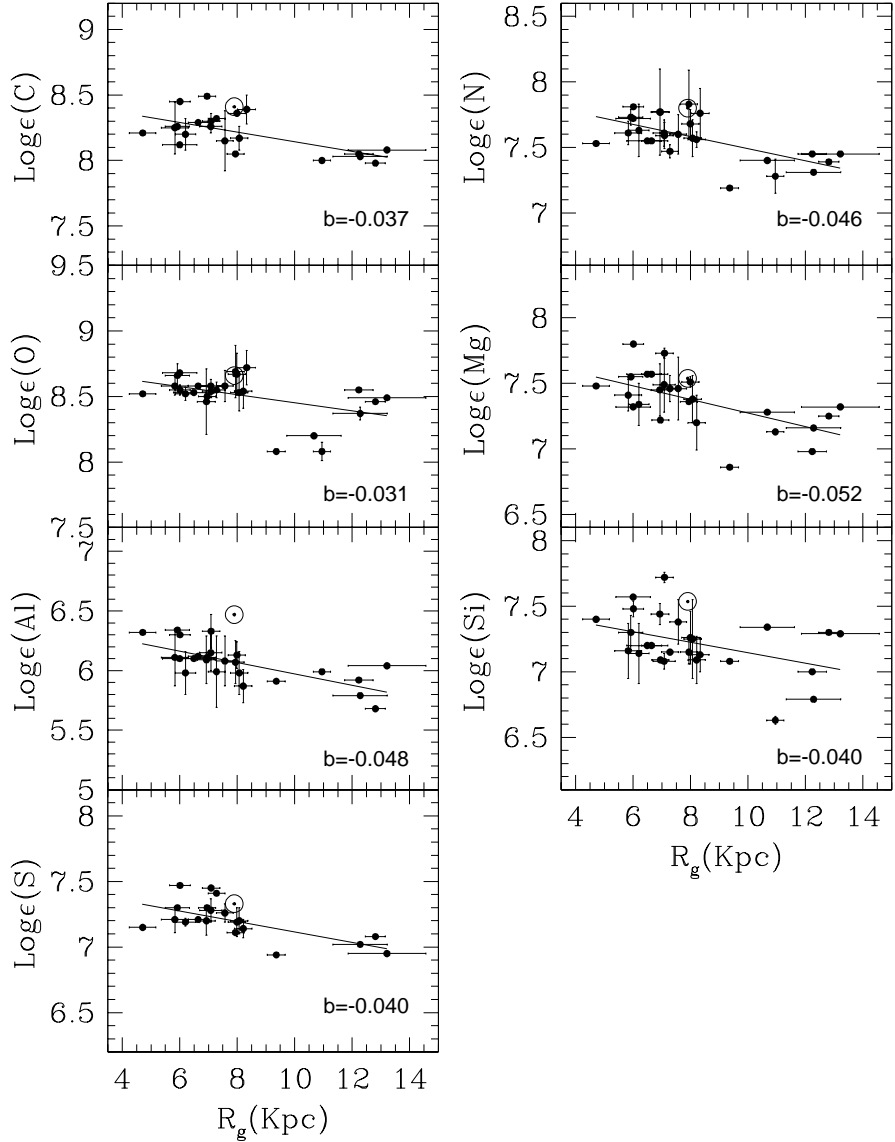}
\caption{Abundance gradients in the 
Galactic disk derived for C, N, O, Mg, Al, Si and S. The abundances adopted for
the clusters are average values of individual stellar abundances in our database. 
The Sun is at $R_{\odot}$=7.9 kpc; solar abundances are from
Asplund (2003 - C and N), Asplund et al. (2004 - O), 
Holweger (2001 -  Mg and  Si), and  Grevesse \& Sauval (1998 - Al and S).
\label{fig2}}
\end{figure}

The derived slopes ($b$ values) are negative for all elements and indicate  
that, as expected, the abundances decline with Galactocentric distance 
but the decline is not very steep.
In addition, it seems that the gradients do not vary significantly and can be considered to be
approximately the same for all studied elements: they can be represented by an 
average slope of $-0.042\pm0.007 {\rm \,{dex\, kpc^{-1}}}$. 
In the different panels of Figure~\ref{fig2}, we display 
the abundance gradients obtained for each element. Horizontal 
errorbars represent the adopted uncertainties in $R_g$ and the vertical ones 
the abundance dispersions for each cluster. For comparison, we also show
the most recent solar abundance results at the Galactocentric distance of 7.9 kpc.
It is clear that these solar results fall within the OB-star
abundance distributions obtained for the studied elements at the solar radius;
except for aluminum whose distribution is below solar by roughly 0.3 dex.
 
Meaningful radial gradients can be also computed considering the individual abundances
derived for each sample star.
One of the advantages being that one can then isolate subsamples
of stars in order to investigate the existence of possible systematic effects 
in the derived gradients.
For instance,
a subsample of stars selected to span a very narrow range in effective temperature can
minimize the effects of systematics in the obtained abundances. 
Another possibility
is to segregate the sample in subsamples of sharp-lined and broad- lined stars
in order to check if there are systematic differences in the obtained abundances and 
corresponding gradients. 
In the following, these two cases are discussed.

\paragraph{Subsample of Stars within a Restricted Range in $T_{\rm eff}$:}\label{teff}

Ideally, to minimize the influence of systematic errors, Galactic radial metallicity 
gradients should be    
obtained from samples of stars within a very narrow range in stellar parameters.
However, it is very difficult to assemble a large
sample of sharp-lined early-type stars if a very restricted range in
effective temperature is imposed. Although this was one of our original goals, our final sample
covers quite a large range in effective temperature from
roughly 19\,000 to 34\,000 K; probably different systematic errors are affecting the
abundances in the stars at the ``cool'' and ``hot'' end of our sample.

\begin{figure}
\epsscale{0.75}
\plotone{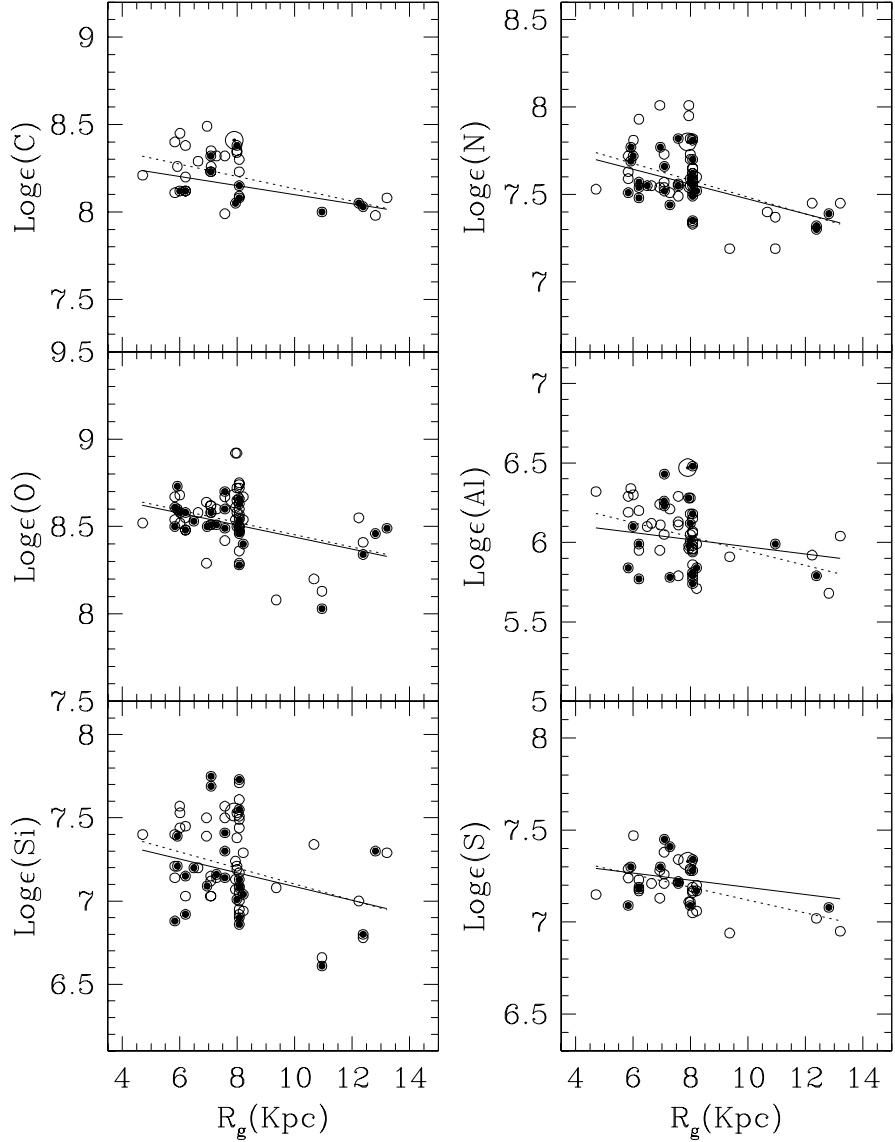}
\caption{A comparison between gradients and stellar abundance distributions for the 
full sample (open circles, dotted lines) and for the subsample 
defined according to the maximum line strength temperature, $T_{\rm max}$ (filled circles, 
solid lines).
From computations of grids of theoretical non-LTE equivalent widths,
$T_{\rm max}$ is reached at approximately: 24\,000~K for C~II; 26\,000~K
for N~II; 27\,500~K for O~II; 25\,000~K for Al~III; 27\,000~K for Si~III;
and 27\,000~K for S~III. The line strengths of the Mg~II triplet at 4481\AA \
reach a maximum at temperatures around 10\,000~K; much lower than the lower limit
for our OB-star sample.
\label{fig3}}
\end{figure}

A subsample defined to have a narrower range in effective temperature was
constructed by adopting as
optimum temperature ($T_{\rm max}$; center of the subsample distribution)
the one for which the line strengths of a given ion reach a maximum.
 In these regimes, the 
abundances are rather insensitive to changes in  $T_{\rm eff}$ and 
in this sense, those stars with temperatures close to $T_{\rm max}$ are expected to present
the most reliable abundance results. As a consequence, gradients derived from subsamples of stars
within a narrow range in $T_{\rm eff}$ around $T_{\rm max}$
are expected to be more robust, even if the ``real'' $T_{\rm eff}$ for the star is
relatively lower or higher than the value adopted in the abundance analysis. 
Given $T_{\rm max}$ for each element (except Mg; Mg~II triplet at 4481\AA \
reach a maximum at temperatures around 10\,000~K) we assembled 
subsamples with stars of effective temperatures within a 2$\sigma$ interval around $T_{\rm max}$; 
where $\sigma$$\sim$4\% is estimated to be the uncertainties in the $T_{\rm eff}$-scale adopted 
in our abundance analyses (see discussion in Paper~I).
The stellar subsamples considered for each studied element are shown as filled circles
in the panels of Figure~\ref{fig3}, while the remaining stars in our sample are represented
by open circles. The corresponding gradients are shown for comparison.
For oxygen, in particular, the similarity of the slopes 
(represented by solid and dotted straight lines) indicates that
the two gradients are practically indistinguishable,
with no indications that significant systematic effects 
are introduced when the full sample, which has a large range in $T_{\rm eff}$, is considered:
the results for oxygen seem therefore
very robust. For all other elements, however, there is a tendency of finding slightly flatter
gradients when subsamples with $T_{\rm eff}$ around $T_{\rm max}$ are adopted. 
These flatter gradients for C, N, O, and Si are all consistent with our original values
within the expected errors. The changes in the gradients of Al and S derived
for the restricted sample in $T_{\rm eff}$ exceed the expected uncertainties.

\paragraph{Subsample of Stars with Low $v \sin i$:}\label{vsini}

Our abundance database includes mostly sharp-lined stars (with relatively 
low $v \sin i$) but, 
because we adopted spectral synthesis technique throughout our abundance analyses,
it also includes 26 OB stars with $v \sin i > 60 {\rm km s}^{-1}$;
all these fast rotators are located within 9 kpc from the Galactic
center and populate the inner part of the Galactic disk. 
The abundances of broad-lined, rapidly rotating stars are expected to be 
somewhat more uncertain than the sharp lined stars, due to  
line blending.
In addition, effects of changes in the abundances due to rotation are more likely
to be influencing the high $v \sin i$ stars, so it is useful to investigate if 
the high $v \sin i$ stars in our sample could be introducing an unexpected bias 
in the derived abundance gradients.

\begin{figure}
\epsscale{0.75}
\plotone{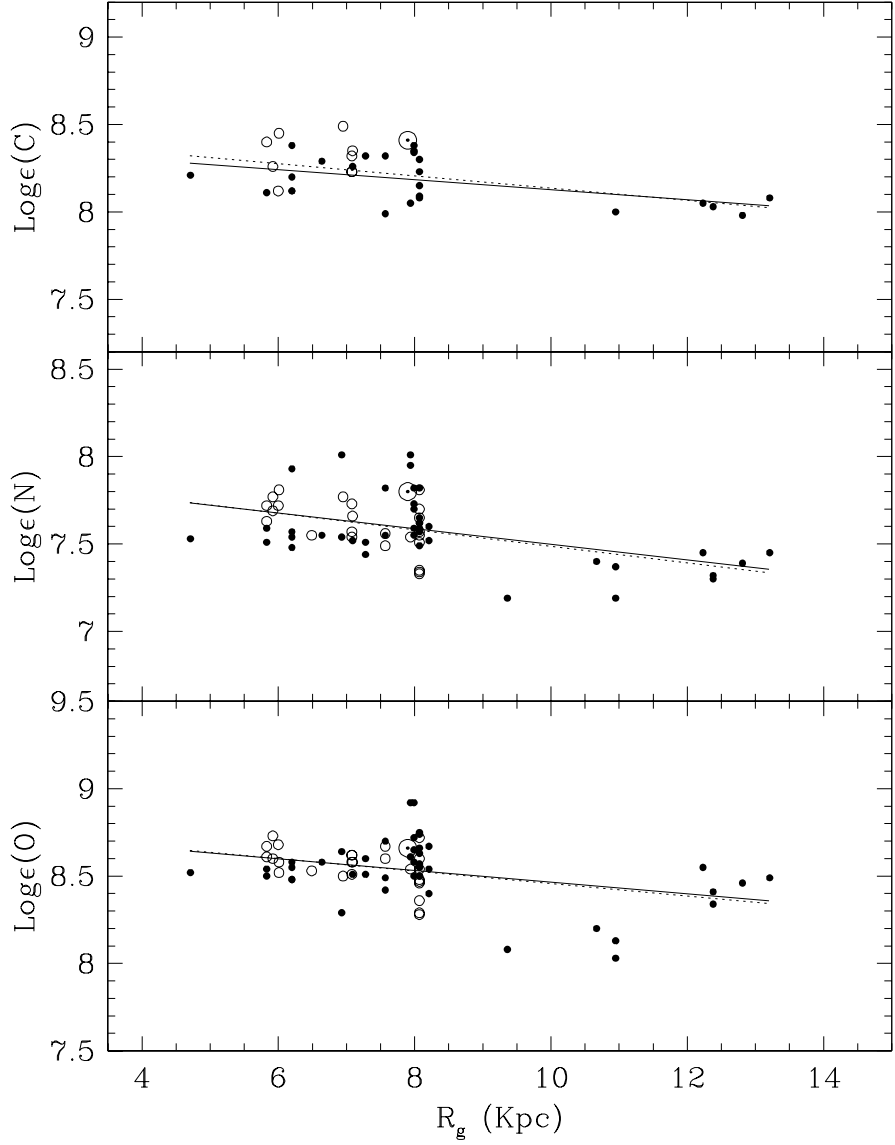}
\caption{A comparison between gradients obtained for the full sample (dotted lines) 
and a subsample of 43 low $v \sin i$ stars (solid lines). The open circles
represent sample stars with high  $v \sin i$ and the filled circles low $v \sin i$ stars.
The inclusion of high $v \sin i$ stars in our analysis does not introduce any significant 
biases in the obtained best-fitted slopes.
\label{fig4}}
\end{figure}

Observational evidence that mixing occurs in our sample
was discussed in Paper~III: 
N overabundances were observed in two of the most massive, most 
evolved, and most rapidly rotating (highest $v \sin i$) members studied in the 
Cep OB2 association. These abundance changes due to mixing raise 
the question of whether the inclusion of high $v \sin i$ stars in 
our sample could be affecting systematically the
C, N and perhaps O abundance gradients obtained. This possibility is investigated here
by segregating our sample in subsamples of stars with low and high $v \sin i$ and 
computing the corresponding gradients.
In the three panels of Figure~\ref{fig4} the sample stars with high $v \sin i$
are represented by open circles, while low $v \sin i$ stars are filled circles.  
We show, for C, N and O, two different abundance gradients:
one obtained for our complete sample (represented by dotted lines) and
another that was computed only for those 43 stars with $v \sin i < 60 {\rm km s}^{-1}$
(represented by solid lines). The conclusion is that the derived slopes are not
significantly different; or that the abundance distribution obtained from analyses of
high $v \sin i$ stars in our sample are in general agreement with what is obtained for
the sharp lined ones. However, 
as we do not know the actual velocity of rotation for the stars, but only
their projected rotational velocity, a fraction of 
the stars that have sharp lines in our sample are also fast rotators seen pole on, so
these two populations are actually mixed in the sharp lined sample.

\subsection{Gradients from the Sample Binned in $R_g$}\label{bin}

Our database of OB stars from open clusters, OB associations and H~II
regions plus Orion, covers 26 disk positions within a range in Galactocentric distance
between 4.7 and 13.2 kpc. The targets, however, are not homogeneously distributed along
the disk. As can be seen from the distribution of abundance  points in Figure~\ref{fig2},
our sample is heavily weighted towards the inner disk and more scarcely sampled outwards. 
Fourteen clusters have $R_g<R_{\odot}$ and lie inside the circle described by the Solar orbit.
Five OB associations (Cep~OB2, Cep~OB3, Cyg~OB7, Ori~OB1, and Lac~OB1) lie very close
to the Sun and are considered also as part of the inner disk.
On the other hand, the chemical distribution of the outer disk is
represented by abundances of only 9 stars that are members of 7 clusters.
 
\begin{figure}
\epsscale{0.75}
\plotone{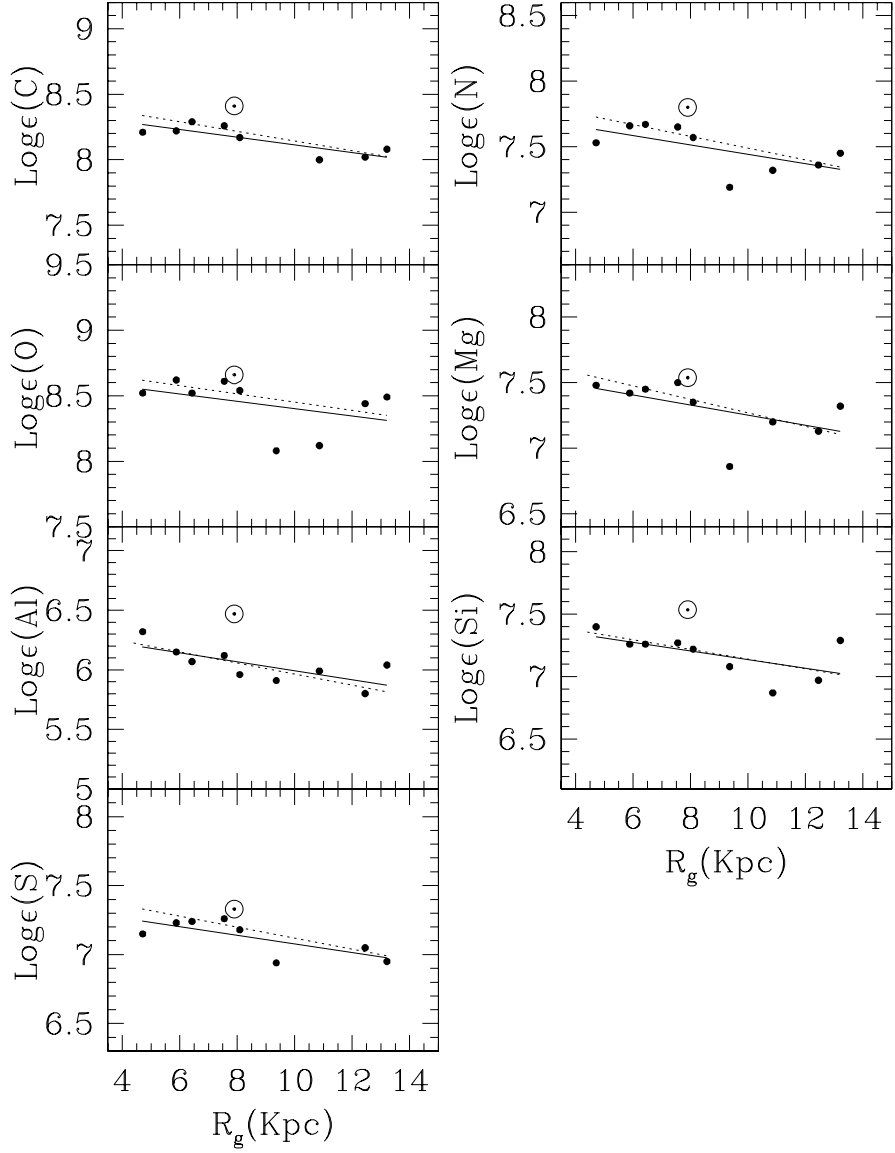}
\caption{Abundance gradients for the sample binned in $\Delta R_g$=1 kpc (solid lines). 
The filled circles represent average abundance values for the stars within a bin. The 
dotted straight lines represent the original gradients presented in Table~\ref{tab3}.
\label{fig5}}
\end{figure}

In order to test if a distribution homogeneuously sampled would result in significantly
different gradients, we divided our sample in 9 bins of 1 kpc  each
(between $R_g$= 4 and 14 kpc) and calculated average abundances
and Galactocentric distances considering those targets available in each bin.
In Figure~\ref{fig5} we represent, for each bin, the average abundances and distances
as solid circles.
New abundance gradients were then calculated for this binned sample and these
are shown as solid straight lines in the figure. For comparison,
we also show the original gradients (derived previously and
based on average abundances per cluster) as straight dotted lines.
A comparison of the best-fitted lines in each panel of Figure~\ref{fig5} indicates that
the new gradients are not significantly
different from the original ones, but we note, that there is a tendency of
finding slightly flatter gradients when the sample is binned.
The new C, N, Mg, Al, and S gradients are shallower than the original gradients
with typical differences of $-0.01 {\rm  dex \, kpc}^{-1}$; the
changes in the gradients of O and Si are smaller than 
$-0.005 {\rm dex \,kpc}^{-1}$.

\section{Discussion}\label{disc}

The derived radial gradients from our database of C, N, O, Mg, Al, Si and S 
abundances are summarized in Table~\ref{tab3}. The main result being that the metallicity
gradients in the Milky Way disk are relatively flat, with an average slope of 
$-0.042 {\rm  dex \, kpc}^{-1}$.
All our efforts in the sense of trying to minimize systematic errors (by binning the sample and
isolating subsamples of stars for which it is expected that the derived abundances
would be most reliable) led to gradients that were, in all cases, flatter 
than the results in Table~\ref{tab3}.
Therefore, there seems to be significant support that, from this sample of
OB stars, the radial metallicity gradients are not steep.
In particular, these flatter gradients are generally in line with the fact that the Milky Way
seems to be a spiral galaxy having a bar of $\sim$3.5 kpc in its Galactic center 
\citep{wes99,beg02}. The observed gradients among spirals show a correlation with
the presence and length of a central bar: there is a tendency of finding steeper gradients 
in normal spiral galaxies when compared to barred spirals 
\citep{mer94}, as the bar would act as to 
promote the homogenization of the Galactic chemical composition, especially in the
region of the inner disk. 

Some interesting abundance patterns seem to emerge from the trends of
abundance with Galactocentric distances presented in this study.
It seems that a few objects in our sample have
abundances that deviate from the best-fitted slopes, being significantly
lower than average by an amount larger than the expected uncertainties.
In the following, we evaluate how much the elemental gradients would
change if one simply rejected from our sample those clusters whose abundances
are the lowest, and most discrepant in each case.

For nitrogen, magnesium and perhaps sulfur, the association of 
Mon~OB2, at a Galactocentric distance of $\sim$ 9.4 kpc,
has lower abundances than expected given its Galactocentric
distance. The metallicity of this OB association, however, is represented by the
abundance of only one star, HD~46202 (analyzed in Paper~VI). Exclusion of
this low abundance star would not significantly change the obtained gradients:
the new gradients without Mon OB2 are  $-0.032$, $-0.046$, and
$-0.033 {\rm  dex \, kpc}^{-1}$, for N, Mg and S, repectively.
For silicon, the abundance of Monnoceros OB2 is not discrepant. 
The lowest Si abundances, however, are found for the two stars members of the open
cluster NGC~2414. When these are discarded,
the Si gradient is flattened to $-0.032 {\rm  dex \, kpc}^{-1}$.

For oxygen, the situation seems more intriging as there are
four stars in our sample with significantly lower oxygen
abundances, all of them lying between 9.4 and 10.9 kpc away from the Galactic center:
HD~46202 (from Mon~OB2), the two targets in NGC~2414, as well as the exciting star of
the H~II region Sh2~247. These objects belong to the only three
clusters/associations
in our sample that fall between $R_g$= 9 and 11 kpc, and they have an average oxygen
abundance of 8.11$\pm$0.07, or, 0.55 dex lower than the solar value.
Our tests indicate that if HD~46202 and
Sh2~247-1 are discarded one at a time, the oxygen gradient remains aproximately unchanged.
However, if we discard the stars in NGC~2414, the gradient flattens to
$-0.024 {\rm  dex \, kpc}^{-1}$. This simple exercise shows that these low
abundance stars in NGC~2414 are playing an important role in making the oxygen
gradient steeper. It should be stressed here, however, that the abundance results for 
these two stars seem quite solid, they
both give consistently low values, and we see no reason why their
abundances should be more uncertain.  In fact, their effective
temperatures ($T_{eff}$ = 23\,260 and 28\,140~K) are such that their abundances
should be considered more robust, since the latter are quite insensitive to errors in
the $T_{eff}$. (We note that these two stars were 
included in the oxygen subsample defined around $T_{\rm max}$ and discussed in
Section~\ref{grads}.)

As dicussed previously, the derived gradients are also subject to uncertainties in 
the adopted distances to the target clusters, and these were taken into account in the 
determination of the best-fitted slopes. Here, in order to investigate what gradient results  
would be obtained if different sets of distances from the literature  were selected
for given clusters, we recalculated the oxygen gradients in a few cases.  
We concentrated on those objects for which there were the largest discrepancies in the
distances found in the literature: we tested all different distance estimates available for 
NGC~6204, NGC~6604,  Vul~OB1, NGC~4755, and Cep~OB2 and verified that the
gradients do not change significantly. We also adopted different distances for 
NGC~1893  and the H~II regions Sh2~247 and Sh2~285. For these, if, for example, 
the smallest distance estimates of 3.6 kpc, 2.2 kpc  and  4.3 kpc, respectively,
were adopted, the oxygen gradient would be
$-0.034 {\rm \,{dex\, kpc^{-1}}}$.  On the other hand, the oxygen gradient
would flatten to  $-0.028 {\rm \,{dex\, kpc^{-1}}}$ if we had considered them to be further away
from the Sun having d=6.0, 3.5 kpc  and 6.9 kpc, respectively. Although these two slopes are
different, it is reassuring that they are within the estimated uncertainty for the oxygen 
gradient quoted in Table 3 ($\sigma b$=0.012 ${\rm \,{dex\, kpc^{-1}}}$).

\subsection{Comparisons with Published Gradients}\label{other}

A summary of the metallicity gradients obtained by recent studies of 
early-type stars starts with the work by  
\citet{gum98}. They analyzed 16 stars belonging to 11
clusters or OB associations with Galactocentric distances between
$5.6< R_g< 13.5$ kpc: ten stars with $R_g >$8~kpc 
are from the sample of Kaufer et al. (1994, who originally found an almost 
null gradient for nitrogen and oxygen); plus six additional stars from 
the inner disk ($R_g <$8~kpc).
Non-LTE abundances were derived self-consistently and their resulting gradients are: 
$-0.035\pm 0.014 {\rm \,{dex\, kpc^{-1}}}$ for carbon,   
$-0.078\pm 0.023 {\rm \,{dex\, kpc^{-1}}}$ for nitrogen,
$-0.067\pm 0.024 {\rm \,{dex\, kpc^{-1}}}$ for oxygen, 
$-0.082\pm 0.026 {\rm \,{dex\, kpc^{-1}}}$ for magnesium, 
$-0.045\pm 0.023 {\rm \,{dex\, kpc^{-1}}}$ for aluminum and
$-0.107\pm 0.028 {\rm \,{dex\, kpc^{-1}}}$ for silicon.
In general, these gradients are steeper than the slopes of abundance with
Galactocentric distances derived in this study (listed in Table~\ref{tab3}); although,
for carbon and aluminum the gradients in both studies are comparable.  

A few comments can be made about the \citet{gum98} sample and
derived gradients. The latter are quite dependent on the
star Sh2~217-3, which lies  $\sim$13 kpc from the  Galactic center.
For oxygen, in particular, if this star is discarded and the oxygen gradient is 
recomputed for their fifteen remaining stars, a shallower slope (by a factor of two) of
$-0.035\pm 0.024 {\rm \,{dex\, kpc^{-1}}}$ is derived. Such a dependance 
on the abundance of one star is understandable because their sample is relatively small and 
more weighted towards smaller Galactocentric distances (fifteen out of sixteen targets have
 $R_g <$ 12 kpc). Five stars in their sample belong to  
the Orion OB1 association, which is located 500 parsecs away from the Sun in the direction
of the anticenter. The oxygen abundance results in \citet{gum98} for the Orion targets
span a range of $\Delta (O)$=0.63 dex (taken to be the 
difference between the largest and smallest oxygen abundance).
It is interesting to note that a similar range in oxygen abundance is found when considering
Gummersbach et al.'s whole sample, that covers a large interval in Galactocentric positions:
$\Delta (O)$ = 0.59 dex. If these oxygen abundance results are in fact representative of 
the abundance distribution of OB stars in the Galactic disk, one possible
interpretation is that the Orion association was able to produce, internally,
the same observed range in oxygen abundances as that encompassed by the radial
distribution along roughly 7 kpc of the Galactic disk.
This is supported by the discussion in \citet{cel94} who observed
a similar spread in oxygen abundances for Orion that could not be easily explained
from the expected uncertainties in the analysis and suggested that a process
of self-enrichment by nucleosynthesis products of Supernovae type II might have
happened in Orion in short timescales. 
 
A large abundance database of OB stars within Galactocentric distances $R_g$=6.1 and 
13.2 kpc has been built over the years 1983 -- 1994 by the Irish group. 
Their abundance data were later extended to larger Galactocentric radii by 
\citet{sma96a,sma96b,sma96c}. This stellar sample, combined with the former, 
was studied by \citet{ser97} who derived oxygen abundances in non-LTE from tabulated theoretical 
equivalent widths of O~II lines published by \citet{beb88}. As discussed in
\citet{cel94}, these published non-LTE calculations used inadequately blanketed model atmospheres 
by \citet{gol84}.
More recently, the observational data (stellar  parameters, distances and equivalent widths)
from the Irish group were compiled by \citet{rol00}.  
They  recomputed the abundances in LTE and  
obtained an oxygen gradient of $-0.067\pm 0.008 {\rm \,{dex\, kpc^{-1}}}$,
for 72 stars at 22 disk positions between $R_g$=6.06 and 17.6 kpc.
Their derived gradients for C, N, Mg, Al, and Si were all roughly similar in magnitude: 
the average gradient for the six elements
analyzed being $-0.068\pm 0.013 {\rm \,{dex\, kpc^{-1}}}$. Their 
gradients are considerably steeper than the slopes obtained here from our
sample. It can be argued, however, that given the 
differences in the techniques adopted in the original
studies from which \citet{rol00} collected their measurements (equivalent widths for different
line sets, and stellar parameters derived non-homogeneously), 
it is possible that different systematic errors (e.g. in the ${T_{\rm eff}}-$scale,
microturbulence, and line sets) may be affecting their absolute abundances, as well
as the final gradients. Moreover, as recognized in 
\citet{rol00}, the adoption of a fully self-consistent non-LTE
approach would result in abundances that are expected to be more accurate on an absolute scale,
even if they conclude in their study that the LTE approach will not influence the 
metallicity gradients themselves.
 
The inner parts of the disk were studied by \citet{sma01}.  
They analyzed four B stars within 2.5--5 kpc of  the 
Galactic center and found LTE oxygen abundances close to solar  
(considering our adopted value of 8.66) for two of them, 
while the other two stars showed overabundances of oxygen relative to the Sun 
of roughly 0.4 dex. These abundance results do not seem consistent with
what would be expected from a simple  
extrapolation, towards smaller Galactocentric distances, of the steep oxygen 
gradient obtained by \citet{rol00}, which would lead to oxygen abundances around
9.2--9.3 dex. On the other hand, however, for C, N, Mg, Al, and Si,
they find enhanced abundances that are roughly in agreement with an extrapolation of 
steep metallicity gradients.
 
As discussed before, the most internal and external parts of the Galactic disk are 
not probed in our study. An extension of our sample would be desirable and 
is possible with the inclusion of additional targets from other literature studies.
In order to investigate what effects the addition of these targets would have
on our derived gradients, we attempted to analyze them using the same
methodology adopted for our sample stars (although not via 
a fully self-consistent analysis because we do not have the observed spectra).
The abundances were not directly taken from the literature,
but were analyzed as follows:  we used the Q-calibration
in order to derive new effective temperatures for
the stars and adopted their surface gravities  from the literature.
We then  calculated non-LTE profiles of O~II lines that reproduced their
published equivalent widths. The Galactocentric distances were
also recalculated for $R_\odot=7.9$kpc, using the mean distances
from the literature, when more than one distance determination
was available. We reanalyzed 10 stars, at 9 new positions in the disk:
star Sh2~217-3, the farthest star and with the lowest abundance in the sample of
\citet{gum98}; the most distant stars in \citet{rol00}, namely
RLWT~13, RLWT~41, Sh2~208-6, Sh2~289-2, Sh2~289-4 and Sh2~283-2;
and three stars of \citet{sma01}, located in the inner parts of the Galactic disk.
These stars were then combined with our sample,
yielding a new gradient of $-0.045\pm 0.010 {\rm \,{dex\, kpc^{-1}}}$ for
oxygen, in the range 2.7--16.0 kpc.
The change in the slope marginally exceeds the expected errors, however, it is
not large enough to make it comparable to a steeper gradient
of  $\sim -0.07 {\rm \,dex\, kpc^{-1}}$. This suggests that
our flatter gradients are not the result of the sample of OB stars that were analyzed.

\subsection{Gradients from Galactic Chemical Evolution Models}\label{gce}

Radial gradients are important large-scale observational constraints
for models of chemical evolution of the Galaxy. 
The disk metallicity gradients in this study were inferred
from the photospheric abundances of young OB stars, assuming that these
abundances represent the composition of the gas from which they formed.

\subsubsection{Are the Abundances of OB Stars Representative of Their Natal Clouds?}\label{natal}

Before comparing our results for OB stars with chemical evolution models
that describe the evolution of the gas with Galactocentric distance, we will discuss briefly 
if it seems reasonable to use abundances in OB stars to trace the metallicity
distribution of the Galactic disk.
This assumption has been questioned in \citet{sem01}, based in part
on the fact that there was a discrepancy between the abundances of OB stars and
the higher solar abundances used as comparison at that time.
To solve this discrepancy, they follow \citet{sno00} and suggest a 
mechanism to reject heavy elements
during star formation, producing the lower heavy-element B-star abundances.
Such mechanisms as proposed by \citet{sno00} are processes of sedimentation 
of refractory elements onto dust grains (and thus decoupled from the gas) and/or
ambipolar diffusion of charged dust grains by magnetic fields.
The situation has changed with the most recent, and lower, solar  
abundance determinations for C, N, and O from 3-D model atmospheres by \citet{asp03}
and \citet{asp04} 
that bring the abundances of OB stars and the Sun into a general agreement
(see review by Herrero 2003). In particular, for our abundance database
(as pointed out in Paper VI) the revised solar oxygen abundance falls roughly at the
peak of the abundance distribution of our 60 sample OB stars from the inner Galactic disk
(see histogram in Figure 2 of Paper~VI),
while the abundances of outer disk B stars are on average lower.

However, even if abundance discrepancies between B stars and
the Sun are probably resolved,
the physical processes that may affect the abundances in the atmospheres of these
hot stars need to be investigated in detail. The indications that
diffusion processes actually take place in these
stellar atmospheres seem weakened. Quiet atmospheres are a requirement for
diffusion to occur. However,
OB stars are usually fast rotators (even stars with
low $v \sin i$ may be fast-rotators seen pole-on); in addition, the
microturbulent velocities derived from non-LTE analysis of OB stars are
typically non-zero and sometimes found to be around 10 ${\rm km\,s^{-1}}$ \citep{gel92,mat02}. 
Therefore, it is expected that  
the velocity fields produced by rotation and microturbulence in the outer
layers of the atmospheres of OB stars may lead to mixing and thus inhibit
diffusion \citep{met02,heh03}.

\subsubsection{Comparisons with Chemical Evolution Models}\label{model}

\begin{figure}
\epsscale{0.75}
\plotone{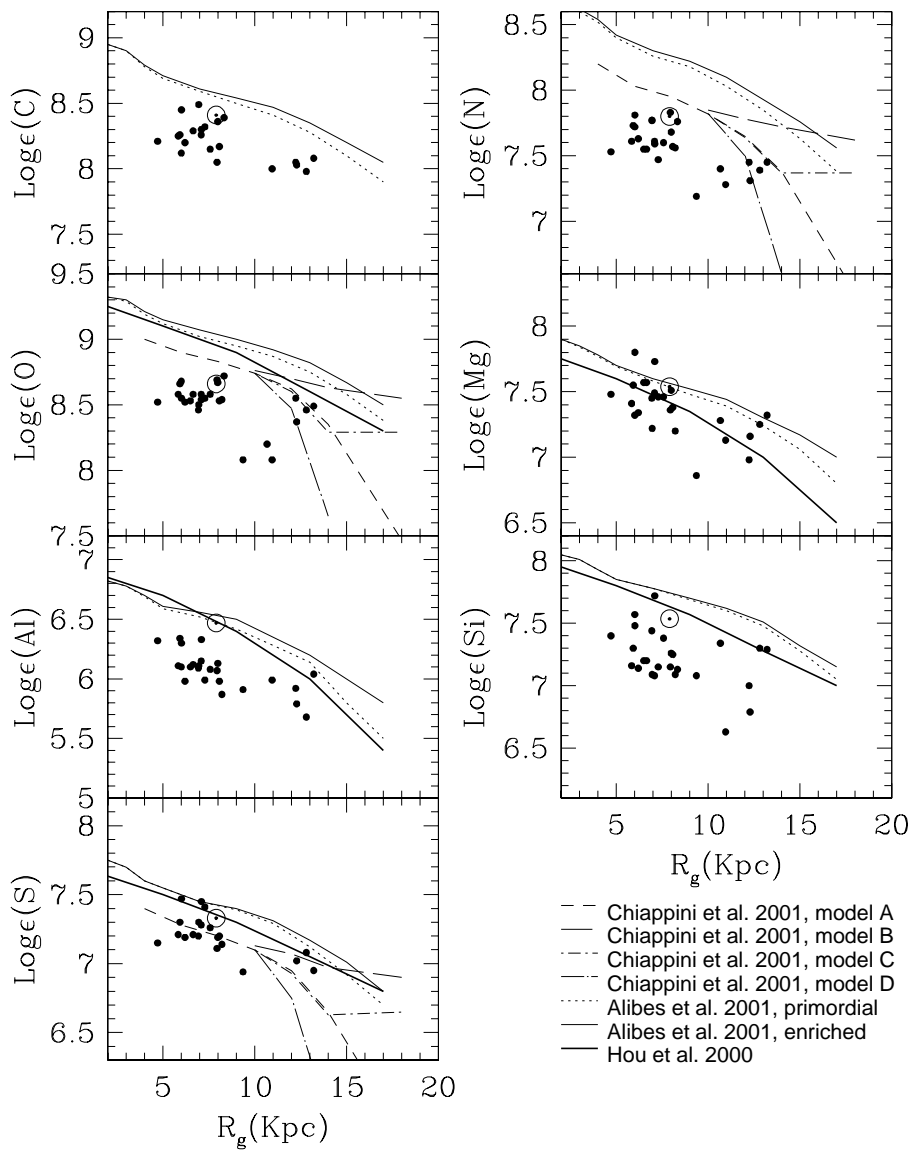}
\caption{Clusters abundances compared to the predicted radial
gradients of Hou, Prantzos, \& Boissier (2000 - solid thick line), 
Alib\'es, Labay, \& Canal (2001 - solid and dotted lines) and 
Chiappini, Matteucci,  \& Romano (2001 - dashed lines).    
\label{fig6}}
\end{figure}

Several Galactic chemical evolution models are available in the literature and these
seek to reproduce observed trends of abundance with Galactocentric distance
in the Milky Way disk.
In Figure~\ref{fig6} we show the abundances obtained for our studied clusters and OB 
associations compared to curves that represent the theoretical radial gradients  
predicted by three different models that assume inside-out formation of the disk, 
i.e., assume differing timescales  for inner and outer disk
formation, with the inner disk  having formed first.

\citet{hpb00} assume infall of external material and  
a multi-slope power-law IMF, with no density threshold for the star formation in the disk.
In addition, they deliberately neglected the yields from intermediate mass stars in 
their models, which led them to conclude that most of abundance distributions, except 
for C and N (not shown in Figure 6), could be reproduced using only the yields of  massive stars.
The model of \citet{ali01b} is an extension of their previous chemical evolution model for 
the solar neighbourhood \citep{ali01a} and also assumes a multi-slope power-law IMF and infall, 
with two different compositions for the infalling matter: primordial and enriched.  
The radial gradients calculated for enriched infalling matter are slightly flatter than those
predicted assuming primodial infalling material: the oxygen gradients are $-0.047$
and  $-0.053{\rm \, dex\,kpc^{-1}}$, respectively. These gradients are 
steeper than the oxygen gradient obtained from our sample. The gradients calculated
for the other elements are also steeper than ours and agree only marginally
considering the uncertainties. 
\citet{chi01} considered four different scenarios in their models of Galactic chemical 
evolution,  by basically changing the halo evolution. One of their models 
(model B) gives the flattest slopes, with the oxygen gradient 
being in  good agreement with the gradient derived in this study. 
On the other hand, steeper gradients, consistent with 
a gradient of $-0.07{\rm \, dex\,kpc^{-1}}$, can be obtained with 
their models A and C (models A and C are similar except that model C assumes
no threshold in the gas density during the halo/thick disk-phase).
The differences between the models A, B, and C (the adopted  halo mass density profiles 
and the density thresholds in the halo phase) suggest that  the differences
in the history of the halo evolution affect mainly the 
outer gradients ($R_g>10$ kpc) whereas the inner gradients are  unchanged.

A simple inspection of the C, N and O panels of Figure~\ref{fig6} 
indicates that it seems that all sets of models attempted to reproduce previous  
(and higher) solar abundances; 
the adoption now of lower solar abundances of CNO produces the observed 
discrepancies between the models and the Sun.
[We recall that we consider the lower values of \citet{asp03,asp04} for CNO]. 
The same applies to the comparison between the models and abundances of OB stars; 
it seems clear that the predicted gradients for
C, N, O, Al and Si fall above the observed abundances, especially in the inner parts 
of the disk. For oxygen, in particular, it is interesting to point out again the significantly
lower abundances obtained for all three studied objects in our sample with
Galactocentric distances between roughly $R_g$= 9 and 11 kpc. This apparent drop in the oxygen 
abundance (if real) is not mimicked by any model; even if we were allowed
to simply slide all the model curves downwards in order to fit the Sun. 
The predicted gradients for Mg are also higher than the observed abundances, but
the discrepancy, in this case, is less pronounced. 
For sulfur, however, there seems to be good agreement between the observations 
(derived abundances) with model B from \citet{chi01}.  

\subsection{Time Evolution of the Abundance Gradients}

The variation of the radial gradients as a function of time is predicted by 
model calculations and also is suggested from observations.
The models by \citet{hpb00} 
and \citet{ali01b} predict a flattening of the radial gradients with time.
Observational indication that the gradients flatten with
time was presented in \citet{fri02} and also in \citet{chw03}, based on analyses of
open cluster metallicities. In these studies, the gradients obtained from clusters in
their youngest age bracket  were always flatter than those derived
for the older clusters. \citet{mcu03} obtained essentially the
same results using planetary nebulae having progenitor stars with different
masses and ages. 

The flatenning of the gradients is generally measured over
Gyr timescales, whereas our sample of young OB stars 
-- 19 out of 26 clusters have ages below 10 Myr \citep{dia02} -- represents in 
principle the present  time composition of the Galactic disk; they 
span a relatively narrow age interval  and are not best suited
to evaluate time variations of the radial gradients. 
However, if we simply segregate from our sample those clusters younger than 10 Myr,
we obtain gradients that are fully consistent with those gradients derived for the whole
sample and listed in Table~\ref{tab3}.
The seven clusters of our sample older than 10 Myr are concentrated in the region between 
6-8 Kpc away from the Galactic center. A meaningful gradient cannot, therefore, be 
computed for this older sample. 

\begin{figure}
\epsscale{0.75}
\plotone{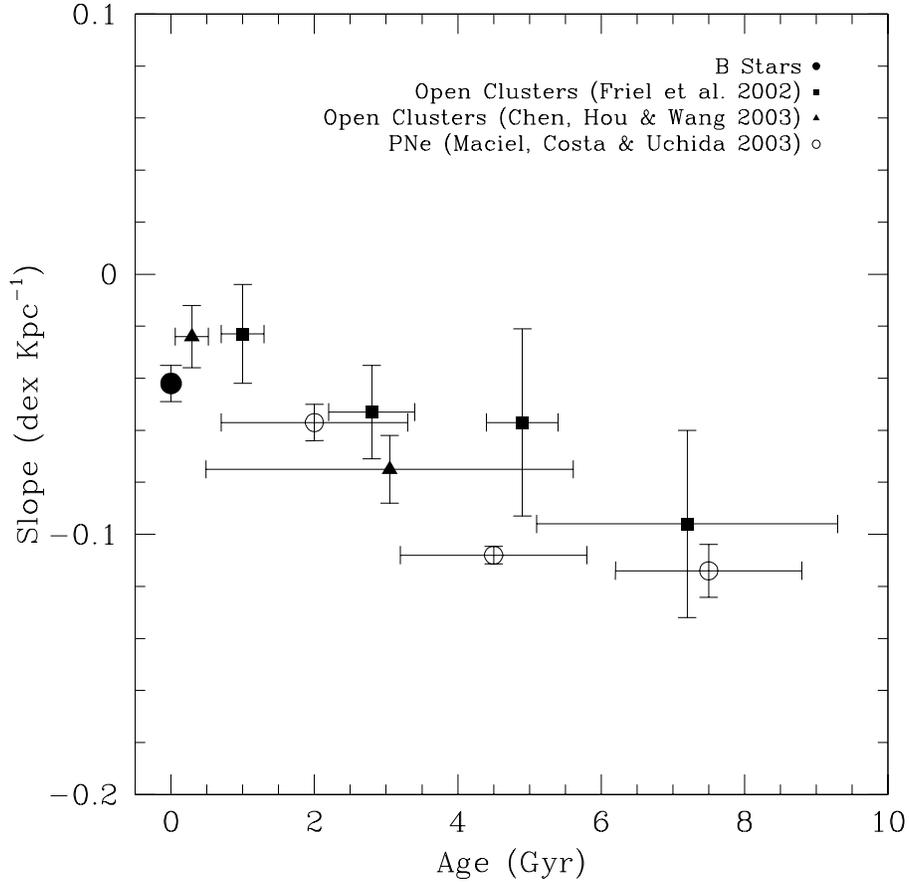}
\caption{The time evolution of the metallicity gradients for the Milky Way disk is shown
for [Fe/H] gradients obtained from open clusters and planetary nebulae. The average abundance
gradient obtained from our sample of OB stars, which are young, is consistent
with the flatenning of the metallicity gradients. 
\label{fig7}}
\end{figure}

It is of interest, however, to compare the present-day gradients we derived from young OB stars
with the time evolution of metallicity gradients inferred from observations of objects 
that span a larger age interval.
This is shown in Figure~\ref{fig7}, where we plot our average slope of 
$-0.042 {\rm \, dex\, kpc^{-1}}$,  
representing the present-day radial metallicity gradient, and 
the metallicity gradients inferred from open clusters and PNe 
with different ages. 
The  clusters [Fe/H] gradients are from \citet{fri02,chw03}
and the age bins are represented in the figure by a mean age plus dispersion
(horizontal errorbars), computed directly from their published data.
We also show the evolution of the [Fe/H] gradients inferred from the oxygen 
gradients of \citet{mcu03},  adopting the [Fe/H]$\times$[O/H] relation presented in their paper. 
For the PNe ages, usually subject to large uncertainties, we adopted simply the 
middle value of each age bin (0--4, 4--5 and $>$ 5 Gyr) and the errorbars represent 
the typical uncertainties assigned to their adopted ages.
An inspection of this figure indicates that the flatter gradients derived for OB stars
in this study are quite consistent 
with a general flatenning of the radial gradients with time.

\section{Conclusions}\label{conc}
 
We have studied the distribution of metallicity with Galactocentric distance for a large sample 
of main-sequence OB stellar members of clusters and associations in the Galactic disk. 
A uniform methodology was adopted  resulting in a homogeneous database
of non-LTE abundances of the elements carbon, nitrogen, oxygen, magnesium,
aluminum, silicon and sulfur that can be used to constrain chemical evolution models.
The radial gradients derived from this abundance database
are flatter than the recent results for early-B stars.
The slope averaged from all elements is found to be $-0.042\pm0.007 {\rm \,dex\, kpc^{-1}}$ and
the oxygen gradient, in particular, is $-0.031 {\rm \,dex\, kpc^{-1}}$.

These flatter gradients are in agreement with the most recent nebular
abundances from H~II regions and PNe that find flatter oxygen gradients for the Galactic disk
\citep{deh00,pil03,hkb04} and are in line with the expected flat gradient caused from
the presence of a bar in the Galactic center \citep{beg02,wes99}. Furthermore, an extension of
the obtained metallicity gradients to the Galactic center is consistent with the 
mean metallicity derived  by \citet{ram00} for a sample of evolved stars 
located within 2.5 pc from the Galactic center, [Fe/H]=$+0.12\pm0.22$. 
Near solar oxygen abundances were also found by \citet{sma01} and
\citet{mun04} for a sample of early B-stars in the innermost Galactic disk,
favouring a flat oxygen gradient towards the Galactic center.
 Our average slope is also consistent with a flattening of the radial gradients as a 
function of time, predicted by models and also supported by observational evidences 
based on abundances in open clusters and PNe.
The derived gradients are believed to be a solid representation of
the behavior of abundance with Galactocentric distances within 5 to 13 kpc
from the Galactic center. 
 
\acknowledgments
We would like to thank Dr. Cristina Chiappini for nice discussions.


\clearpage
\begin{deluxetable}{ccccccccc}
\tabletypesize{\scriptsize}
\tablecaption{Cluster Abundances  \label{tab1}}
\tablewidth{0pt}
\tablehead{
\colhead{Object} &
\colhead{$\log \epsilon$(C)[n]} & 
\colhead{$\log \epsilon$(N)[n]} & 
\colhead{$\log \epsilon$(O)[n]} &
\colhead{$\log \epsilon$(Mg)[n]} &
\colhead{$\log \epsilon$(Al)[n]} & 
\colhead{$\log \epsilon$(Si)[n]} & 
\colhead{$\log \epsilon$(S)[n]}}
\startdata
Sh2 47  & 8.21[1] & 7.53[1] & 8.52[1] & 7.48[1] & 6.32[1] & 7.40[1] & 7.15[1] \\
NGC 6611& $8.25\pm0.20$[2] & $7.61\pm0.09$[4] & $8.58\pm0.07$[4] & $7.41\pm0.12$[4] & $6.11\pm0.24$[3] & $7.16\pm0.21$[4] & $7.21\pm0.10$[3] \\
Sh2 32  & 8.26[1] & $7.73\pm0.06$[2] & $8.66\pm0.09$[2] & 7.55[1] & 6.34[1] & $7.30\pm0.13$[2] & 7.30[1] \\
NGC 6204& 8.12[1] & 7.72[1] & 8.68[1] & 7.32[1] & 6.10[1] & 7.57[1] & \nodata \\
Trumpler 27 & 8.45[1] & 7.81[1] & $8.55\pm0.04$[2] & 7.80[1] & 6.30[1] & $7.48\pm0.06$[2] & 7.47[1] \\
NGC 6231& $8.20\pm0.12$[4] & $7.63\pm0.20$[4] & $8.52\pm0.05$[4] & $7.34\pm0.16$[3] & $5.98\pm0.18$[4] & $7.14\pm0.23$[4] & $7.19\pm0.03$[4] \\
NGC 6604& \nodata  & 7.55[1] & 8.53[1] & 7.57[1] & 6.10[1] & 7.20[1] & \nodata \\
Ara OB1 & 8.29[1] & 7.55[1] & 8.58[1] & 7.57[1] & 6.12[1] & 7.20[1] & 7.21[1] \\
Vul OB1 & \nodata  & $7.77\pm0.33$[2] & $8.46\pm0.25$[2] & $7.45\pm0.20$[2] & $6.09\pm0.20$[2] & $7.44\pm0.08$[2] & $7.20\pm0.11$[2] \\
Stock 16& 8.49[1] & 7.77[1] & 8.50[1] & 7.22[1] & 6.11[1] & 7.09[1] & 7.30[1] \\
Sct OB2 & $8.30\pm0.06$[2] & $7.59\pm0.10$[2] & $8.54\pm0.05$[2] & $7.73\pm0.04$[2] & $6.33\pm0.14$[2] & $7.72\pm0.04$[2] & 7.45[1] \\
NGC 4755& $8.26\pm0.05$[3] & $7.61\pm0.10$[3] & $8.58\pm0.05$[4] & $7.49\pm0.21$[4] & $6.15\pm0.15$[2] & $7.08\pm0.06$[4] & $7.28\pm0.09$[3] \\
IC 2944 & 8.32[1] & $7.47\pm0.05$[2] & $8.55\pm0.06$[2] & $7.46\pm0.10$[2] & $5.99\pm0.30$[2] & $7.15\pm0.01$[2] & 7.41[1] \\
Cyg OB3 & $8.15\pm0.23$[2] & $7.60\pm0.15$[4] & $8.58\pm0.12$[5] & $7.46\pm0.24$[4] & $6.08\pm0.21$[4] & $7.38\pm0.17$[5] & $7.26\pm0.07$[3] \\
Cyg OB7 & 8.05[1] & $7.83\pm0.26$[3] & $8.69\pm0.20$[3] & $7.36\pm0.02$[2] & $6.07\pm0.18$[2] & $7.15\pm0.09$[3] & 7.11[1] \\
Lac OB1 & $8.36\pm0.02$[3] & $7.68\pm0.11$[5] & $8.67\pm0.16$[5] & $7.51\pm0.04$[3] & $6.13\pm0.11$[4] & $7.26\pm0.20$[5] & $7.19\pm0.11$[4] \\
Cep OB2 & $8.17\pm0.09$[5] & $7.57\pm0.14$[17]& $8.53\pm0.14$[17]& $7.38\pm0.18$[16]& $5.98\pm0.18$[16]& $7.25\pm0.30$[17]& $7.20\pm0.10$[6] \\
Cep OB3 & \nodata   & $7.56\pm0.06$[2] & $8.54\pm0.13$[3] & $7.20\pm0.21$[3] & $5.87\pm0.14$[3] & $7.09\pm0.18$[3] & $7.14\pm0.07$[3] \\
Ori OB1*& $8.39\pm0.11$[15] & $7.76\pm0.13$[15] & $8.72\pm0.13$[18] & \nodata & \nodata & $7.13\pm0.13$[18] & \nodata \\
Mon OB2 & \nodata   & 7.19[1] & 8.08[1] & 6.86[1] & 5.91[1] & 7.08[1] & 6.94[1] \\
Sh2 247 &  \nodata  & 7.40[1] & 8.20[1] & 7.28[1] &  \nodata & 7.34[1] &  \nodata \\
NGC 2414& 8.00[1] & $7.28\pm0.13$[2] & $8.08\pm0.07$[2] & 7.13[1] & 5.99[1] & $6.63\pm0.03$[2] &  \nodata \\
Sh2 253 & 8.05[1] & 7.45[1] & 8.55[1] & 6.98[1] & 5.92[1] & 7.00[1] &  \nodata \\
NGC 1893& 8.03[1] &$7.31\pm0.01$[2] & $8.37\pm0.05$[2] & 7.16[1] & 5.79[1] & $6.79\pm0.01$[2] & 7.02[1] \\
Sh2 284 & 7.98[1] & 7.39[1] & 8.46[1] & 7.25[1] & 5.68[1] & 7.30[1] & 7.08[1] \\
Sh2 285 & 8.08[1] & 7.45[1] & 8.49[1] & 7.32[1] & 6.04[1] & 7.29[1] & 6.95[1] \\
\enddata
\tablenotetext{*}{non-LTE abundances from Cunha \& Lambert (1994).}
\end{deluxetable}

\clearpage
\begin{deluxetable}{crrlcc}
\tabletypesize{\scriptsize}
\tablecaption{Cluster Distances  \label{tab2}}
\tablewidth{0pt}
\tablehead{
\colhead{Cluster} &
\colhead{$l(^\circ)$} &
\colhead{\,\,\,\,\,$b(^\circ)$} &
\colhead{d (kpc)} & 
\colhead{$\langle{\rm d}\rangle$ (kpc)} & 
\colhead{${\rm R}_g$ (kpc)}}
\startdata
Sh2 47       &  15.3 &    0.1 & $3.7^1, 3.1^2 $                      & 3.4$\pm$0.4 & 4.7$\pm$0.5 \\
NGC 6611     &  17.0 &    0.8 & $2.19^3, 2.5^4, 2.6^5,  1.68^6 $     & 2.2$\pm$0.4 & 5.8$\pm$0.5 \\
Sh2 32       &   7.3 & $-$2.0 & $1.8^8, 2.2^9$                       & 2.0$\pm$0.3 & 5.9$\pm$0.4 \\
NGC 6204     & 338.3 & $-$1.1 & $2.51^3, 2.6^4, 1.32^6, 1.94^7 $     & 2.1$\pm$0.6 & 6.0$\pm$0.6 \\
Trumpler 27  & 355.1 & $-$0.7 & $2.0^5, 1.65^{11}, 2.1^{32} $        & 1.9$\pm$0.2 & 6.0$\pm$0.3 \\
NGC 6231     & 343.5 &    1.2 & $1.8^4, 1.6^5, 1.77^6,  2.0^{10}$    & 1.8$\pm$0.2 & 6.2$\pm$0.3 \\
NGC 6604     &  18.3 &    1.7 & $0.70^4, 2.1^5, 1.64^{12}$           & 1.5$\pm$0.7 & 6.5$\pm$0.7 \\
Ara OB1      & 336.3 & $-$1.4 & $1.38^3, 1.4^6, 1.59^7,  1.1^{13} $  & 1.4$\pm$0.2 & 6.6$\pm$0.3 \\
Vul OB1      &  59.4 & $-$0.1 & $2.0^3, 2.54^7, 3.5^{17}$            & 2.7$\pm$0.7 & 6.9$\pm$0.3 \\
Stock 16     & 306.1 &    0.1 & $1.9^5, 2.0^{14} $                   & 1.9$\pm$0.1 & 6.9$\pm$0.3 \\
Sct OB2      &  39.0 &    7.6 & $1.0^3, 1.17^{15}  $                 & 1.1$\pm$0.1 & 7.1$\pm$0.3 \\
NGC 4755     & 303.2 &    2.5 & $2.34^4, 1.03^6, 1.9^{16}$           & 1.8$\pm$0.7 & 7.1$\pm$0.4 \\
IC 2944      & 294.6 & $-$1.4 & $2.1^4, 2.0^5, 1.95^6, 2.2^{18} $    & 2.1$\pm$0.1 & 7.3$\pm$0.3 \\
Cyg OB3      &  73.5 &    2.0 & $2.29^3, 1.9^5,   2.31^7   $         & 2.2$\pm$0.2 & 7.6$\pm$0.3 \\
Cyg OB7      &  90.0 &    2.0 & $0.83^3,0.79^{21}$                   & 0.8$\pm$0.1 & 7.9$\pm$0.3 \\
Lac OB1      &  96.8 & $-$16.1& $0.6^3, 0.63^7, 0.368^{19}$          & 0.6$\pm$0.1 & 8.0$\pm$0.3 \\
Cep OB2      &  99.2 &    3.8 & $0.83^3, 0.95^5, 0.96^7, 0.615^{19}$ & 0.8$\pm$0.2 & 8.1$\pm$0.3 \\
Cep OB3      & 110.4 &    2.8 & $0.87^3, 0.725^6, 0.84^7  $          & 0.8$\pm$0.1 & 8.2$\pm$0.3 \\
Ori OB1      & 205.0 & $-$17.4& $0.5^3, 0.43^4, 0.56^7, 0.438^{19}$  & 0.5$\pm$0.1 & 8.3$\pm$0.3 \\ 
Mon OB2      & 206.5 & $-$2.1 & $1.51^3, 1.62^6, 1.63^7, 1.67^{20} $ & 1.6$\pm$0.1 & 9.4$\pm$0.3 \\
Sh2 247      & 188.9 &    0.8 & $3.5^{23}, 2.2^{24}$                 & 2.8$\pm$0.9 & 10.7$\pm$0.9 \\
NGC 2414     & 231.0 &    2.0 & $3.98^{21}, 4.2^{22}$                & 4.1$\pm$0.1 & 10.9$\pm$0.3 \\
Sh2 253/Bo 1 & 192.4 &    3.2 & $4.4^{23}, 4.8^{27}, 4.06^{29}$      & 4.4$\pm$0.4 & 12.2$\pm$0.5 \\
NGC 1893    &173.6&$-$1.7& $4.0^5,3.7^6, 3.6^{25}, 4.3^{26}, 4.8^{27}, 6.02^{28}$ &4.4$\pm$0.9&12.3$\pm$0.9 \\
Sh2 284/Do 25& 211.9 & $-$1.3 & $5.2^{23}, 5.6^{30}, 5.5^{33} $      & 5.4$\pm$0.2 & 12.8$\pm$0.3 \\
Sh2 285      & 213.9 & $-$0.6 & $6.9^{23,30}, 4.3^{31}, 6.4^{24}$    & 5.9$\pm$1.4 & 13.2$\pm$1.3 \\
\enddata
\tablerefs{
\scriptsize{
1: \citet{cra78},  2: \citet{lah85},  3: \citet{hum78}, 
4: \citet{alt70},  5: \citet{fei94},  6: \citet{bef71},  
7: \citet{mee95},  8: \citet{bli82},  9: \citet{vem75},
10: \citet{cea71}, 11: \citet{van80}, 12: \citet{mev75a}, 
13: \citet{keg92}, 14: \citet{cra71}, 15: \citet{rei90}, 
16: \citet{sho84}, 17: \citet{sej81}, 18: \citet{tov98},  
19: \citet{dez99}, 20: \citet{per87}, 21: \citet{hem84}, 
22: \citet{fem80}, 23: \citet{mof79}, 24: \citet{lah87}, 
25: \citet{cuf73}, 26: \citet{tap91}, 27: \citet{fit93}, 
28: \citet{mar00}, 29: \citet{mev75b}, 30: \citet{tem93}, 
31: \citet{rol94}, 32: \citet{mof77}, 33: \citet{len90} 
}}
\end{deluxetable}

\clearpage
\begin{deluxetable}{ccccccc}
\tabletypesize{\scriptsize}
\tablecaption{Radial Gradients of Elemental Abundances ($a + bx$)  \label{tab3}}
\tablewidth{0pt}
\tablehead{
\colhead{$\log$(X/H) } & 
\colhead{$a$} & 
\colhead{$\sigma a$} &
\colhead{$b$} &
\colhead{$\sigma b$} & 
\colhead{n} & 
\colhead{R}}
\startdata
C  &  8.513 & 0.086 &  -0.037 & 0.010 & 21 & -0.64 \\
N  &  7.950 & 0.093 &  -0.046 & 0.011 & 26 & -0.65 \\
O  &  8.762 & 0.105 &  -0.031 & 0.012 & 26 & -0.46 \\
Mg &  7.794 & 0.121 &  -0.052 & 0.014 & 25 & -0.61 \\
Al &  6.452 & 0.085 &  -0.048 & 0.010 & 24 & -0.71 \\
Si &  7.546 & 0.147 &  -0.040 & 0.017 & 26 & -0.43 \\
S  &  7.514 & 0.092 &  -0.040 & 0.011 & 20 & -0.64 \\
\enddata
\end{deluxetable}


\clearpage

\clearpage

\clearpage

\clearpage

\clearpage

\clearpage

\clearpage

\end{document}